\documentstyle[prl,aps]{revtex}

\begin{document}
\input psfig
\pssilent
\title{Inspiralling black holes: the close limit}

\author{Gaurav Khanna$^1$, John Baker$^{1,2}$, Reinaldo J. Gleiser$^3$,
Pablo
Laguna$^4$,\\ Carlos O. Nicasio$^4$, Hans-Peter Nollert$^5$, Richard
Price$^6$, Jorge Pullin$^1$}
\address{1. Center for Gravitational Physics and Geometry,
Department of Physics\\
The Pennsylvania State University, 104 Davey Lab, University Park PA 16802}
\address{2. Max-Planck-Institut f\"ur Gravitationsphysik,
        Albert-Einstein-Institut,\\
        Schlaatzweg 1,
        D-14473 Potsdam,
        Germany}
\address{3. Facultad de Matem\'atica, Astronom\'{\i}a y F\'{\i}sica,
Universidad Nacional de C\'ordoba,\\
Ciudad Universitaria, 5000 Cordoba, Argentina}
\address{4. Department of Astronomy and Astrophysics,
The Pennsylvania State University, \\525 Davey Lab, University Park, PA
16802}
\address{5. Theoretische Astrophysik, Universit\"at T\"ubingen,
72076, T\"ubingen, Germany}
\address{6. Department of Physics, University of Utah, Salt Lake City, UT
84112}

\maketitle

\begin{abstract}
Using several approximations, we calculate an estimate of the gravitational
radiation emitted when two equal mass black holes coalesce at the end of
their binary inspiral. We find that about 1\% of the mass energy of the
pair will emerge as gravitational waves during the final ringdown and
a negligible fraction of the angular momentum will be radiated.
\end{abstract}
\vspace{-9cm} 
\begin{flushright}
\baselineskip=15pt
CGPG-99/5-3  \\
gr-qc/9905081\\
\end{flushright}
\vspace{7cm}
\pacs{4.30+x}

An international network of interferometric gravitational wave (GW)
observatories (the LIGO project in the US, the VIRGO
and GEO projects in Europe and the TAMA
project in Japan \cite{gw}) will be capable of detecting
gravitational waves in the next few years. This could have
revolutionary implications for astronomy since it constitutes a new
form of ``light'' with which to observe the universe, a form that is
better correlated with the bulk motions of matter and is very
hard to shield or distort. Gravitational waves were originally
predicted by Einstein in 1915 but their theoretical existence was not
completely understood until the 60's \cite{Kennefick} and their direct
experimental detection has remained a daunting challenge
\cite{Saulson}. One of the most promising sources for detection is the
collision of two black holes to form a single, final hole.  Such a
collision is expected to be the end point of the decaying inspiral for
a binary pair of holes.  Just how promising such collisions are depends
very much on the masses of the colliding holes, and is connected with
the characteristics of detectors.  The sensitivity of
the laser interferometric detectors, like LIGO, peaks around 100 Hz.

The major determinant of the frequency of GWs produced in a black hole
inspiral/collision is $M$, the mass of the system (and of the final
black hole formed). The frequency scales as $1/M$, and is around
$10^{4}$\,Hz for a $1M_{\odot}$ system. The usual black hole
candidates are stellar mass holes, and supermassive holes
($\sim10^{6}M_{\odot}$) in the centers of many galaxies. The
frequency of GWs from the former candidates would be too high except
for the weak radiation from the early inspiral. GWs from supermassive
holes would be too low in frequency for Earth based systems, but well
suited to space based detection.  Black holes of $100M_{\odot}$
would be ideal as sources and black holes of any mass can in principle
exist. But the astrophysical motivation for such ``middleweight''
holes was weak. Recent observations \cite{middleweight} of x-ray
emission from galaxies suggest that some galaxies may contain
middleweight holes, perhaps as an evolutionary stage in the formation
of supermassive holes. Should such a middleweight hole exist it will
produce GWs near the optimal $\sim100$\,Hz frequency when it collides
with compact objects of equal or smaller mass.

For two different reasons, we focus here on the collision of roughly
equal mass holes. The first is that the power radiated from the
collision of holes of masses $m_{1}\leq m_{2}$ scales as
$(m_{1}/m_{2})^{2}$, at least in the case that $m_{1}$ is
significantly smaller than $m_{2}$. The GW strain produced in a
detector will then scale as $m_{1}/m_{2}$, and collisions of equal
mass holes will be the most detectable. The second reason is that
computations in the case $m_{1}\ll m_{2}$ can be done by treating the
smaller mass as a particle perturbing the spacetime of the hole. By
contrast, even a qualitative understanding is lacking for the
collisions of equal mass holes at the endpoint of binary inspiral.  It
is not known, for example, whether the inspiralling holes will
smoothly and gradually merge to form a single final hole, or whether
the smooth inspiral will end with a plunge of the holes into each
other.  It is commonly accepted that numerical solutions of Einstein's
field equations, on supercomputers, will be a necessary element in
understanding such collisions.  Although numerical relativity has
given a fairly complete picture of head-on collisions, numerical
studies of the astrophysically interesting case, truly three
dimensional collisions, are proving to be remarkably
difficult. Progress with this problem is characterized by approximate
and imperfect results.  We present here such a result: the first
approximate calculation for the gravitational radiation emitted by the
collision at the end of the inspiral of two equal mass holes. We find,
in particular, that only about 1\% of the mass of the two hole system
will be radiated in the collision. The several approximations that
limit the generality of this conclusion will be spelled out below.

The radiation of interest is generated
at the end of the slow quasi-Newtonian binary inspiral, when the
interaction between the holes becomes highly nonlinear.  It is useful
to break down the subsequent evolution of the binary into two stages:
the ``merger'' describing the transition from two disjoint holes to a
single final, highly distorted hole; and the ``ringdown'' stage in
which the final hole relaxes to a stationary final state by the
emission of GWs.  Here we will consider only GW radiation during
ringdown.  In particular, we use the close limit method in which we
treat the late stage evolution as a perturbation of the final
hole. The evolution can then be computed with the relatively simple
linear equations of perturbation theory, rather than the nonlinear
equations of Einstein's full theory. To evolve a spacetime in general
relativity, one needs to provide initial data, a 3-geometry $g_{ab}$
and an extrinsic curvature $K_{ab}$, that solve Einstein's equations
on some starting hypersurface (i.e., at some starting time). For two
black holes, this is an easy task if the holes are far apart, since
one can superpose the solutions for two individual holes ignoring
their interactions. When the black holes are close on the initial
hypersurface, the astrophysically correct initial data is the solution
corresponding to what would have evolved during the binary inspiral,
but such an evolution cannot be computed with present day  techniques.  One
must therefore use a somewhat artificial initial data solution that is
a best guess at a representation of close black holes.  The need for
such a guess is one of the sources of uncertainty in our result. It
should be noted that the choice of initial data is not a shortcoming
if our answers are to be used as code checks for numerical relativity,
since the same initial data can be used for our close limit evolution
and in numerical relativity codes.

A convenient prescription for initial data has been given by Bowen and
York \cite{BoYo} and is used in much of numerical relativity.  Their
method assumes that the initial spatial geometry metric $g_{ab}$ is
conformal to a flat space according to $g_{ab} = \phi^4
\delta_{ab}$. One specifies the location of any number of holes in the
conformal flat space and specifies parameters representing the
momentum and spin of each hole. In the limit that the holes are far
from each other on the initial hypersurface, these parameters have the
usual physical meaning.  When the holes are close, the ``separation,''
``spin,'' and ``momentum '' of an individual hole has no unambiguous
meaning.  It would seem {\em a priori}\, that any evolution done with
uncertain initial data is of little value. But a large set of examples
\cite{GlNiPrPuboost} computed for head-on collisions, and compared
with numerical relativity results, gives us a basis for considering
that our predictions, based on ``reasonable'' initial data choices have
a certain degree of validity, at the very least as order of magnitude
estimates, but more likely as correct within a factor of two or so.

In the Bowen-York  formalism the extrinsic curvature for a hole $i$
with no spin, and with momentum $\vec{P}^{(i)}$, is
$K _{ab} =\phi^{-2}\widehat{K} _{ab}$ where
\begin{equation}\label{KBY}
\hat{K}^{(i)}_{ab} = {3
\over 2 r_{(i)}^2} \left[ 2 P^{(i)}_{(a} n^{(i)}_{b)}
-(\delta_{ab}-n^{(i)}_a n^{(i)}_b)P_{(i)}^c n^{(i)}_c\right]\ .
\end{equation}
Here $r_{(i)}$ and $\vec{n}^{(i)}$ are respectively the distance to
hole $i$, and the direction from the hole, in the conformal flat
space. In Eq.~(\ref{KBY}) and below we use units in which $G=c=1$.
Our choice of initial data is a Bowen-York specification that most
plausibly represents a binary pair of nonspinning holes.  In
particular our initial data is taken to represent equal mass holes at
$x=\pm L/2$ in the conformal flat space. The holes are taken to have
momenta $\vec{P}^{(1)}$ and $\vec{P}^{(2)}$ that have equal magnitude
$P$, and that have opposite directions  orthogonal to the line
joining the holes.  The equations satisfied by $\widehat{K} _{ab}$ are
linear and independent of $\phi$, so for the two hole system we can
take the extrinsic curvature to be that given by the sum
$\widehat{K}_{ab}=\widehat{K}^{(1)}_{ab}+\widehat{K}^{(2)}_{ab}$ of
the contributions of each of the holes. It should be noted that
$\widehat{K}_{ab}\rightarrow0$ to first order in $L$ as
$L\rightarrow0$.

The conformal factor $\phi$ is required to satisfy a nonlinear
elliptical equation $\nabla^2 \phi = -K_{ab} K^{ab} /8 \phi^7$ that is
not analytically solvable.
To get a closed form expression for $\phi$, we use the ``slow''
approximation: For small $P$ the product $K_{ab} K^{ab}$ is second
order in $P$ and one can approximate the solution by a solution of
$\nabla^2 \phi=0$.  Solutions of the latter with the topology of two
holes are well known and it is known that this approximation works
better than might be expected \cite{Bakeretal,GlNiPrPuboost}. For
larger values of the momenta, $K_{ab}$ grows linearly whereas $\phi$
grows as a fractional power of the momenta. Therefore for larger
values the extrinsic curvature dominates  the initial data and
the error we are making in the determination of the conformal factor
$\phi$ is again negligible. This was explicitly verified in a study of
the head-on collisions of boosted black holes
\cite{Bakeretal,GlNiPrPuboost}.

With the choice of $\widehat{K}_{ab}$ and approximate solution for
$\phi$ the specification of initial data is complete. In the limit
that $L\rightarrow 0$, the initial data are those of a stationary
hole, so we can view $P, L$ or $PL$ as an expansion parameter, and we can
evolve forward in time using perturbation theory.
It would seem that this perturbative evolution should be done with the
Teukolsky formalism\cite{T} for perturbations of Kerr holes, since the final
hole formed in binary inspiral is expected to be a rapidly rotating
(i.e., Kerr) hole, perhaps possessing angular momentum $J$ that is near the
extreme Kerr limit $J/M^{2}=1$ for black holes.  It turns out,
however, that the initial data described above are not a perturbation
of a Kerr hole. In the space of parameters $P$ and $L$, there is no
family of solutions, with fixed nonzero angular momentum, that has a
Kerr hole as the zero perturbation limit. This can be viewed as a
consequence of the very constraining requirements that must be
satisfied by Kerr initial data\cite{KrPr}.

We {\em can} consider the initial data to be perturbations of a
nonrotating (Schwarzschild) hole.  The initial data are purely even
parity and we use the perturbation formalism developed by
Zerilli\cite{ZERI}, in which even parity perturbations are
decomposed in angular multipoles. As is usually the case, radiation is
dominantly quadrupolar. The two hole data has $\ell=2$ radiation for
$m=0,\pm2$. For example the Zerilli function $\psi_{(2,2)}$,
describing the $\ell=2,m=2$ multipole has the Cauchy data
\begin{eqnarray}
\psi_{(2,2)}(0,r) &=& 4 \sqrt{{2 \pi \over 15}} M L^2 {r\;(7 \sqrt{r}+5
\sqrt{r-2M}) \over (2r+3M) (\sqrt{r}+\sqrt{r-2M})^5}\\
\dot{\psi}_{(2,2)}(0,r)  &=& -{i\sqrt{30 \pi} \over 10} PL
{(4 r +3 M)\sqrt{r-2 M} \over r^{5/2} (2 r +3 M)}
\end{eqnarray}
and the Zerilli function for $(\ell=2,m=-2)$ is the complex conjugate
of $\psi_{(2,2)}(t,r)$

It will be very important to know the range of validity of our
perturbative result, that is, the maximum value of $PL$, or angular
momentum, for which our results are reasonably accurate. This question
is equivalent to the question ``how early in the coalescence can the
close limit approximation be used?''  To answer this, we rely on
experience with computations of head-on collisions, where comparisons
were available with numerical results for fully non-linear evolutions.
A more objective approach would be to compute answers to second-order
in the perturbation, as has been done successfully for head-on
collisions \cite{GlNiPrPuboost}.  Since our present approach is
limited by the use of initial data that would make our predictions
only roughly accurate, the very lengthy computations of second-order
theory do not seem justified at the present time.  Instead, as a
``poor person's'' approach to estimating the error of our
calculations, we have also carried out the evolution of the spacetime
using the Teukolsky formalism. For each choice of the angular momentum
perturbation $PL$, we have taken the initial data to specify a
perturbation of a Kerr spacetime with angular momentum parameter
$a=PL/M$. In this hybrid approach, a change in the parameter $PL$ does
not take us along a family of perturbations of a single spacetime, but
the result of the calculation will be ``first order'' in $PL$ in some
sense. The difference between the result of Teukolsky evolution and
Zerilli evolution, therefore, gives us a crude indicator of the limit
of validity of first-order perturbation theory.
\begin{figure}
\centerline{\psfig{figure=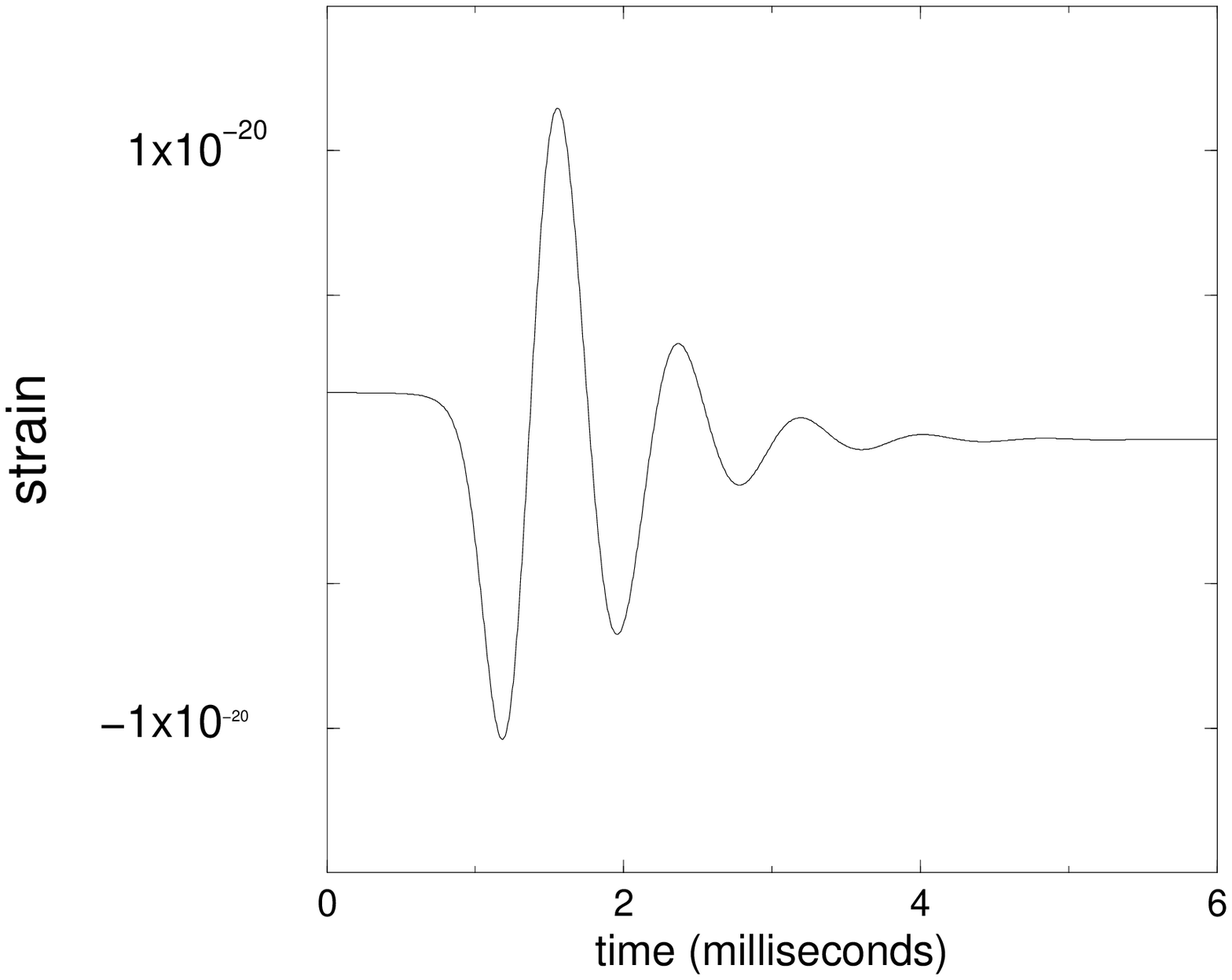,height=60mm}
\psfig{figure=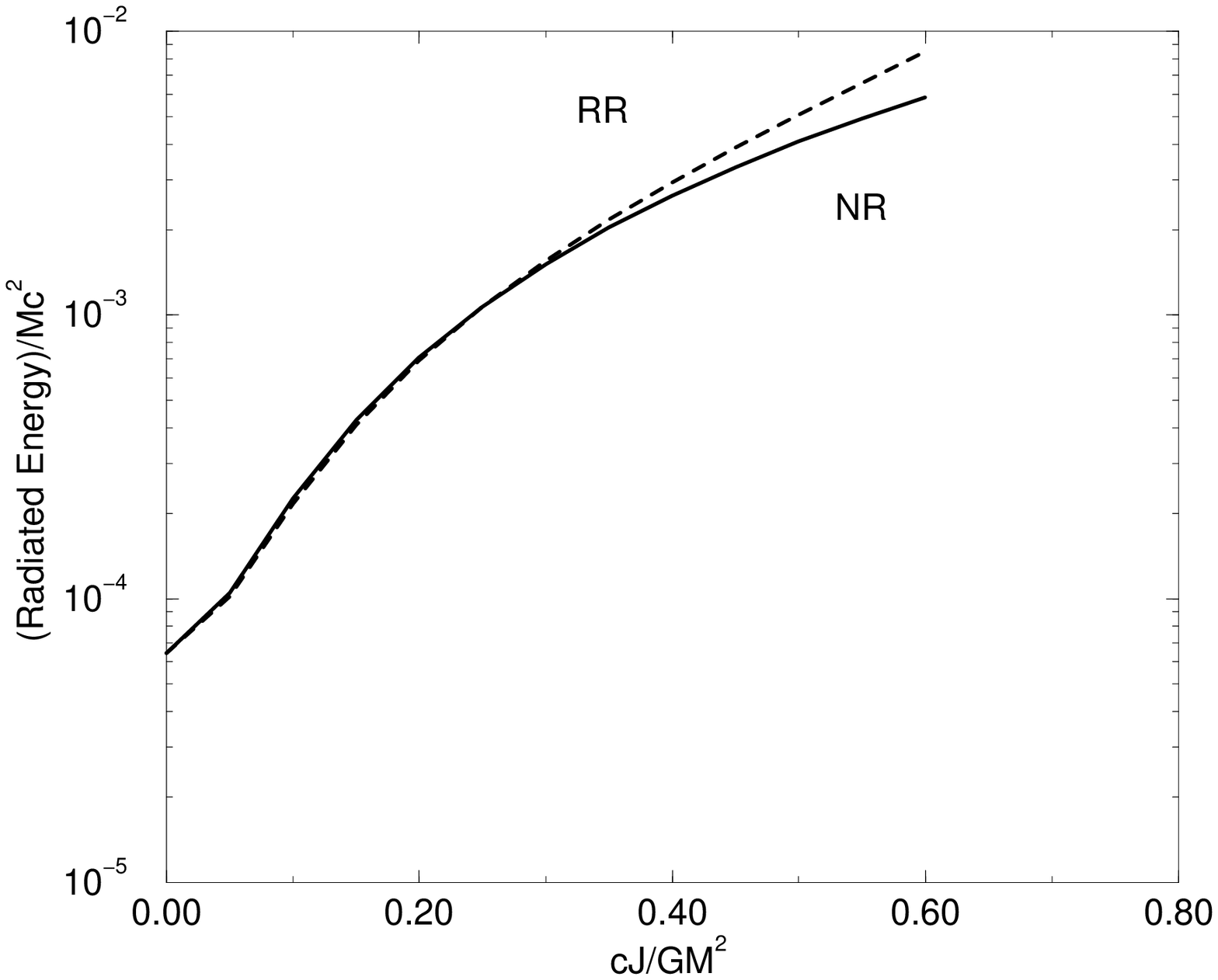,height=60mm}}
\caption{ The figure on the left shows the strain amplitude in the
equatorial plane as a function of time, producted by a $10M_{\odot}$
black hole binary going through its ringdown phase at a distance of
100\,Mpc from the detector.
We assume that the
detector is oriented for maximum sensitivity to the radiation in the
orbital plane. (For an L shaped laser interferometer, one arm
of the L would have to have the orientation just described for
the bar; the other arm would have to be perpendicular to the orbital
plane.)
On the right is shown the fraction
of the mass of the system radiated as gravitational waves as a
function of the normalized initial angular momentum of the collision.
}
\end{figure}

The curve on the left in Fig.~1 shows the waveform of the
gravitational radiation, that is, the strain that would affect the arm
lengths of an interferometer. The curves on the right show the total
energy carried in the waves as a function of the expansion parameter
$\epsilon=J/M^{2}$ .  The result shown is for two black holes
initially separated in conformal flat space by $L=1.8$ in terms of the
mass of each hole. (If one were considering a Misner type geometry,
the proper separation measured along the geodesic threading the
throats would be $5.5$ in in the same units \cite{etaletal}.)  The curve
labeled $Z$ shows results for linearized perturbation calculations
using the Zerilli equation; the curve $T$ shows the result of
``hybrid perturbation'' calculations using the Teukolsky equation.
The two results diverge around parameter values of $\epsilon=0.4$ to
$0.5$, and this is a reasonable limit to take for the applicability of
perturbation estimates. We note that the Teukolsky results lie above
the Zerilli results, and this weakly suggests that the Zerilli-based
estimates are more accurate. (In close limit estimates for head-on
collisions, linearized results always overestimated the nonlinear --- i.e.,
numerical relativity --- results.)

The limitation to $\epsilon$ less than or around $0.5$ was expected
but is unfortunate, since most realistic collisions will probably take
place with angular momentum closer to $\epsilon=1$ \cite{FlHu}.  One
can make an educated guess of what could happen around $\epsilon\sim
1$ by looking at the curves in Fig.~1, and by extrapolating by eye to
$\epsilon\sim 1$.  It is difficult in this way to make a case for
radiation more than $1\%$ of the system's mass. This is in rough
agreement with $3\%$ value used in data analysis calculations of Flanagan
and Hughes\cite{FlHu}.

The radiated angular
momentum is well defined in linear perturbation theory even for spacetimes
without any symmetry. It can be computed in two very different ways, either
from an angular momentum flux constructed from the Landau-Lifshitz
pseudotensor or by looking at the change in the $\ell=1$ odd parity
perturbation. In either case we find that the radiated
angular momentum is given by
\begin{equation}
\Delta J = {3  \over 2 \pi} \lim_{r \rightarrow \infty}
 \int_{0}^{-\infty} \left[ \mbox{Im}(\psi) {\partial \mbox{Re}(\psi)
 \over \partial t} - \mbox{Re}(\psi) {\partial \mbox{Im}(\psi) \over
 \partial t}\right] dt
\end{equation}
where we have written $\psi_{(2,2)}(t,r) = \mbox{Re}(\psi) + i
\mbox{Im}(\psi)$. For values of $\epsilon$ from around 0.1 to 0.5, the
angular momentum radiated is roughly 0.1\% of the initial angular
momentum. For a given multipole with azimuthal index $m$, and single
frequency $\omega$, the angular momentum radiated divided by the energy
radiated will be less or equal than
$|m/\omega|$\cite{FlHu,KST-RMP}. If we take $|\omega|$ to be the real
part of the least damped $\ell=2$ quasinormal mode, and we take $m=2$,
this tells us that the ratio of $E/M$ radiated to $J/M^{2}$ radiated
should be $\sim0.2$, and this is approximately the case near
$\epsilon=0.5$.

In summary, we have used the ``close limit'' to estimate the radiation
in the collision at the end of the inspiral of two equal mass
nonrotating black holes. The assumptions and restrictions were: (i)
only the ``ringdown'' radiation was computed; (ii) we assumed that a
simple initial data set gave an adequate representation of appropriate
astrophysical conditions; (iii) we assumed that the final hole is not
near the extreme Kerr limit; (iv) we used close limit estimates of the
evolution.  Our main conclusion is that the energy radiated in
ringdown is probably not more than 1\% of the total mass of the
system. The most serious uncertainty in this result is the possibility
that the radiation from the early merger stage of coalescence is very
much larger than the ringdown radiation.  With our 1\%$Mc^{2}$
estimate, collisions of black holes of $100M_{\odot}$ would be
detectable with signal to noise of 6 out to distances on the order of
200Mpc by the initial LIGO configuration and to distances of 4Gpc with
the advanced LIGO detector.

This work was supported in part by grants
NSF-INT-9512894,
NSF-PHY-9357219,
NSF-PHY-9423950,
NSF-PHY-9734871,
NSF-PHY-9800973,
PHY-9407194,
NATO-CRG971092, DFG-SFB-382,
by funds of the University of
C\'ordoba, Utah, and Penn State.  We also acknowledge support of CONICET and
CONICOR (Argentina).  JP also acknowledges support from the Alfred P.
Sloan and the John S.  Guggenheim foundations.  RJG is a member of CONICET.


\end{document}